\documentclass[pra,twocolumn,letterpaper,aps,showpacs,fixfloat,superscriptaddress,reprint]{revtex4-1}
\usepackage{graphicx,amsmath,amsfonts,amssymb}

\usepackage[colorlinks, linkcolor=blue, citecolor=blue, urlcolor=blue, breaklinks=true]{hyperref}

\newcommand{\fref}[1]{Fig.~\ref{#1}}
\newcommand{\Fref}[1]{Figure~\ref{#1}}

\newcommand\rmd{\mathrm{d}}
\newcommand\tra{\mathrm{T}}
\renewcommand\vec[1]{\boldsymbol{#1}}
\newcommand\mat[1]{\boldsymbol{#1}}
\newcommand\abs[1]{\lvert#1\rvert}
\newcommand\expt[1]{\langle#1\rangle}
\DeclareMathOperator\ceil{ceil}
\DeclareMathOperator\diag{diag}
\DeclareMathOperator\tr{Tr}

\newcommand{\ket}[1]{\lvert#1\rangle}

\newcommand{\braket}[2]{\langle#1\vert#2\rangle}

\newcommand{\im}[1]{\text{Im}\!\left\{#1\right\}}
\newcommand{\re}[1]{\text{Re}\!\left\{#1\right\}}

\makeatletter
\newlength \figurewidth
\makeatother

\begin{document}

\pacs{42.50.Wk, 07.10.Cm, 07.60.Ly, 42.79.Gn}

\title{Reconfigurable long-range phonon dynamics in optomechanical arrays}

\date{\today}

\author{Andr\'e Xuereb}
\email[Corresponding author. ]{andre.xuereb@um.edu.mt}
\affiliation{Department of Physics, University of Malta, Msida MSD\,2080, Malta}
\affiliation{Centre for Theoretical Atomic, Molecular and Optical Physics, School of Mathematics and Physics, Queen's University Belfast, Belfast BT7\,1NN, United Kingdom}
\author{Claudiu Genes}
\affiliation{Institut f\"ur Theoretische Physik, Universit\"at Innsbruck, Technikerstrasse 25, A-6020 Innsbruck, Austria}
\affiliation{IPCMS (UMR 7504) and ISIS (UMR 7006), Universit\'{e} de Strasbourg and CNRS, Strasbourg, France}
\author{Guido Pupillo}
\affiliation{IPCMS (UMR 7504) and ISIS (UMR 7006), Universit\'{e} de Strasbourg and CNRS, Strasbourg, France}
\author{Mauro Paternostro}
\affiliation{Centre for Theoretical Atomic, Molecular and Optical Physics, School of Mathematics and Physics, Queen's University Belfast, Belfast BT7\,1NN, United Kingdom}
\author{Aur\'elien Dantan}
\affiliation{QUANTOP, Danish National Research Foundation Center for Quantum Optics, Department of Physics and Astronomy, University of Aarhus, 8000 Aarhus C, Denmark}

\begin{abstract}
We investigate periodic optomechanical arrays as reconfigurable platforms for engineering the coupling between multiple mechanical and electromagnetic modes and for exploring many-body phonon dynamics. Exploiting structural resonances in the coupling between light fields and collective motional modes of the array, we show that tunable effective long-range interactions between mechanical modes can be achieved. This paves the way towards the implementation of controlled phononic walks and heat transfer on densely-connected graphs as well as the coherent transfer of excitations between distant elements of optomechanical arrays.
\end{abstract}

\maketitle

\noindent\section{Introduction}
Optomechanical systems (OMS), naturally lying in the intersection between optical technologies and electronics, play a major role in communication and information-processing sciences~\cite{Roukes2001}. Recent advances in the fabrication of high-quality mechanical resonators and their integration with electromagnetic fields have allowed to bring the control of mechanical motion to, or close to, the quantum regime, with potential applications in areas as different as metrology and sensing, quantum information processing, and tests of the fundamental laws of physics~\cite{Kippenberg2008,Meystre2013,Aspelmeyer2013}. While these investigations have principally focused on the interplay between electromagnetic radiation and \emph{single} mechanical resonators, \emph{multi}-element OMS are beginning to be actively studied theoretically~\cite{Eisert2004,Bhattacharya2008,Hartmann2008,Ludwig2010,Dobrindt2010,Stannigel2010,Heinrich2011,Chang2011,Stannigel2012, Xuereb2012c,Seok2012,Ludwig2012b,Xuereb2013,Chesi2014}, as well as experimentally~\cite{Lin2010,Mahboob2011,Mahboob2012,Massel2011,Zhang2012,Botter2013}. The motivations for exploring their potential are manifold. First, their multi-mode nature makes them well suited for applications in communication technology~\cite{Roukes2001,Stannigel2010}. In addition, they hold the promise for enhanced performance in quantum optomechanics and metrology~\cite{Xuereb2012c,Seok2012}. Finally, the common interaction of several mechanical elements with one or more electromagnetic fields allows, in principle, for the engineering of complex long-range interactions among the mechanical components, paving the way to the investigation of quantum many-body phenomena with macroscopic elements~\cite{Ludwig2010,Xuereb2012c,Ludwig2012b,Tomadin2012,Xuereb2013}. A key challenge in OMS is to engineer \emph{reconfigurable} systems, in which the interactions are not predetermined by the bulk properties of the system but can be tailored and switched on or off. This would open the way towards, e.g., efficient and controlled manipulation of heat transfer and single excitations in optomechanical arrays.

In this paper we propose to use periodic optomechanical arrays as reconfigurable platforms for engineering the coupling between multiple mechanical and electromagnetic modes. Such a device operates in a regime where the array is \emph{transmissive} and light permeates through the structure. This allows both for the enhancement of the optomechanical response~\cite{Xuereb2012c} and the coupling to specific collective motional modes of the array~\cite{Xuereb2013}. We show that effective long-range phonon--phonon interactions can be achieved by addressing these transmissive modes. Arising from structural resonances defined by the light fields, these interactions are naturally tunable and reconfigurable. We provide two illustrations of controlled many-body dynamics made possible in this setting: (i)~In the bad-cavity regime of optomechanics, the structure acts as a beamsplitter array for phonons with effective long-range mode coupling, enabling the investigation of phononic random walks on highly-connected graphs and controlled transfer of heat between distant elements in the array. (ii)~In the good-cavity regime, coherent and reconfigurable transfer of single excitations is shown to be possible between distant array elements.\\
These results should enable the investigation of, e.g., non-standard heat transport and thermodynamics as well as excitation and information transfer in a wide range of periodically ordered OMS, e.g., nanoelectromechanical resonators~\cite{Buks2002,Teufel2011}, microtoroids~\cite{Kippenberg2005,Arbabi2011}, dielectric membranes~\cite{Thompson2008} or particles~\cite{RomeroIsart2011}, optomechanical crystals~\cite{Eichenfield2009}, or cold atoms~\cite{Botter2013}. The engineering of genuine quantum many-body effects in such an array of mesoscopic systems will provide an additional element into the ``mechanical quantum simulator'' that we propose here. This will allow for addressing, e.g., fundamental issues related to the persistence of quantum features in multi-element systems with comparatively large masses, dimensions, and at finite temperature. While these conditions would normally imply Newtonian mechanics, the results presented here suggest that clear signatures of non-classical behavior can persist even in such a mesoscopic simulator.

\begin{figure}[t]%
 \includegraphics[width=0.9\figurewidth]{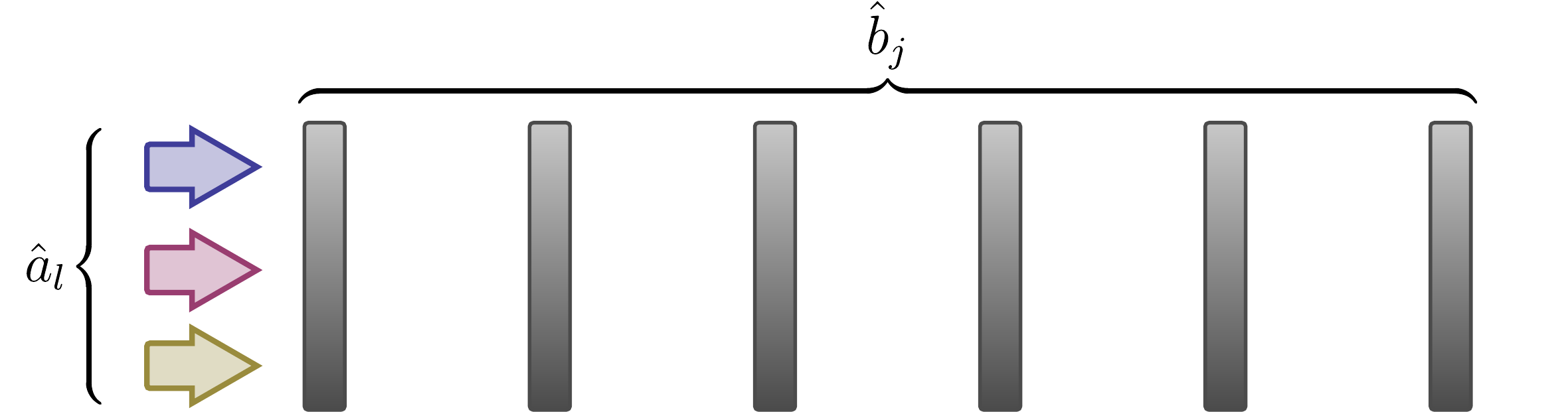}\\[2mm]
 \includegraphics[width=0.9\figurewidth]{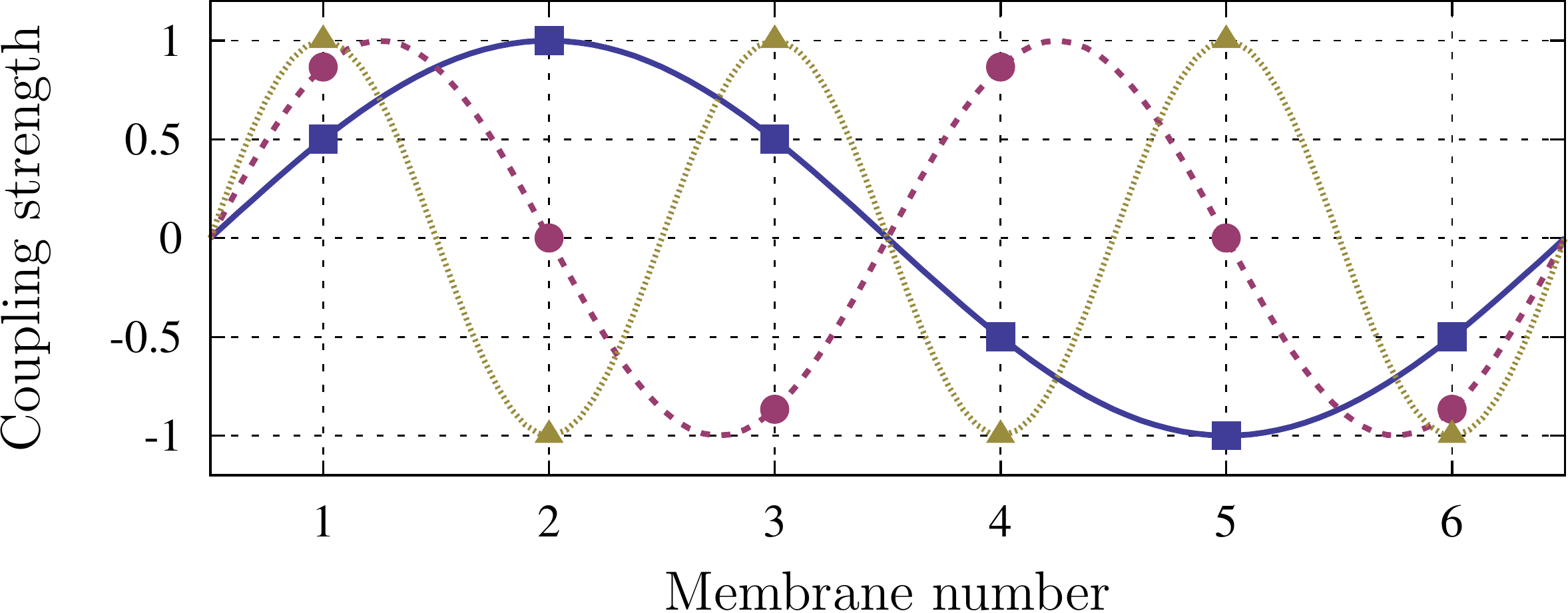}%
 \caption{(Color online) Top: Notation used for referring to the mirrors (index $j$) and light fields ($l$). Bottom: Optomechanical coupling for the `transmissive' modes $l=1$ (squares), $l=2$ (circles), and $l=3$ (triangles) in a $6$-element array.}%
 \label{fig:ModeProfiles}%
\end{figure}%

\noindent\section{Generic model}
We consider the dynamics of an externally driven optomechanical system composed of $N$ identical mechanical elements, here dubbed `mirrors,' and $N-1$ optical cavity modes. The mirrors are modeled as harmonic oscillators with annihilation operators $\hat{b}_j$, vibrational frequency $\omega$, and decay rate $\gamma$. The $l$\textsuperscript{th} optical mode is detuned by $\Delta_l$ with respect to its driving field, has a decay rate $\kappa$, and is described by the annihilation operator $\hat{a}_l$. Here, $j=1,\dots,N$ and $l=1,\dots,N-1$. We treat the mechanical oscillators as a periodic array of lossless 1D scatterers operating in the Lamb--Dicke regime. Such an array displays $N-1$ optical resonances for which the effective reflectivity vanishes~\cite{Deutsch1995,Xuereb2012c} and for which the `transmissive' light modes strongly couple to collective motional modes of the array~\cite{Xuereb2012c,Xuereb2013}.
The Hamiltonian of the system reads~(see Appendix) ($\hbar=1$)
\begin{equation}
\label{eq:Ham}
\hat{H}=\sum_{j,l}\bigl[\omega\hat{b}_j^\dagger\hat{b}_j^{\vphantom{\dagger}}+\Delta_l\hat{a}_l^\dagger\hat{a}_l^{\vphantom{\dagger}}+\epsilon_{l,j}(g_l^{\vphantom{\ast}}\hat{a}_l^\dagger+g_l^\ast\hat{a}_l^{\vphantom{\dagger}})(\hat{b}_j^\dagger+\hat{b}_j^{\vphantom{\dagger}})\bigr]\,,
\end{equation}
where the `coupling vectors' $\vec{\epsilon}_l=(\epsilon_{l,j})_j$ are dimensionless, have unit norm, and are determined mainly by the optical properties of the system. In the case of a periodic array of identical scatterers, at the frequencies where the array is transparent these vectors take the sinusoidal form $\epsilon_{l,j}\varpropto\sin[2\pi l(j-1/2)/N]$~\cite{Xuereb2013}. The optomechanical couplings of the elements thus have a long-ranged sinusoidal profile spanning the whole array (Fig.~\ref{fig:ModeProfiles}). Each complex frequency $g_l\varpropto\alpha_l$ is determined by the mean field amplitude of the respective mode ($\alpha_l$) and the overall optomechanical coupling strength multiplying $\vec{\epsilon}_l$.
\par
Hamiltonian~\eqref{eq:Ham} allows for the engineering of a flexible toolbox for the manipulation of phonon dynamics in an optomechanical array. In the following we shall investigate two regimes: (i)~In the `bad-cavity' regime ($\kappa\gg\omega$) we derive an effective Hamiltonian for the mechanics and investigate phonon diffusion and heat transfer through the array. (ii)~In the `good-cavity' regime ($\kappa\ll\omega$) we derive an analytical expression for the matrix describing the unitary evolution, which allows for the engineering of controlled coherent phonon dynamics.

\begin{figure}[t]%
 \includegraphics[width=0.33\figurewidth]{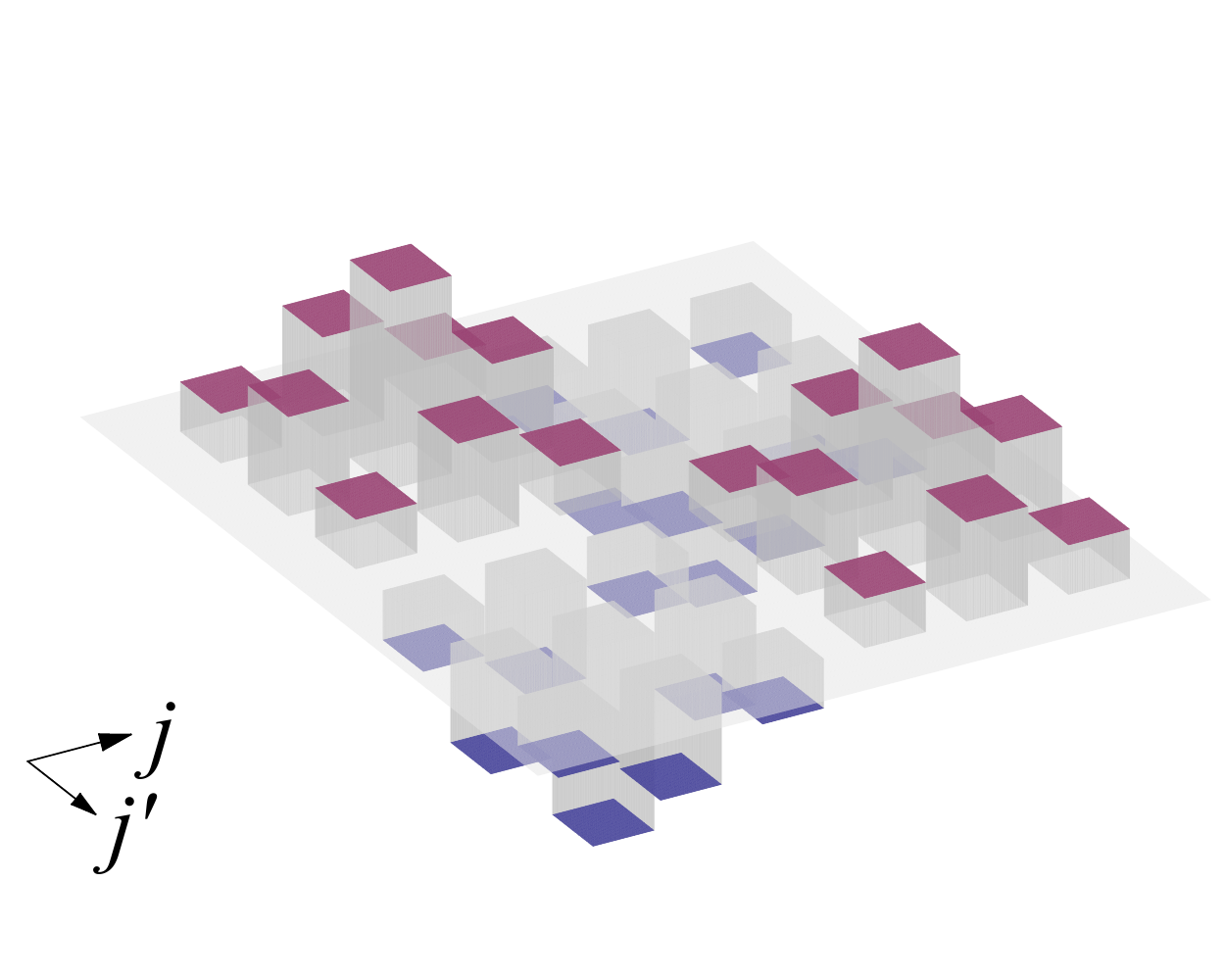}\quad%
 \includegraphics[width=0.33\figurewidth]{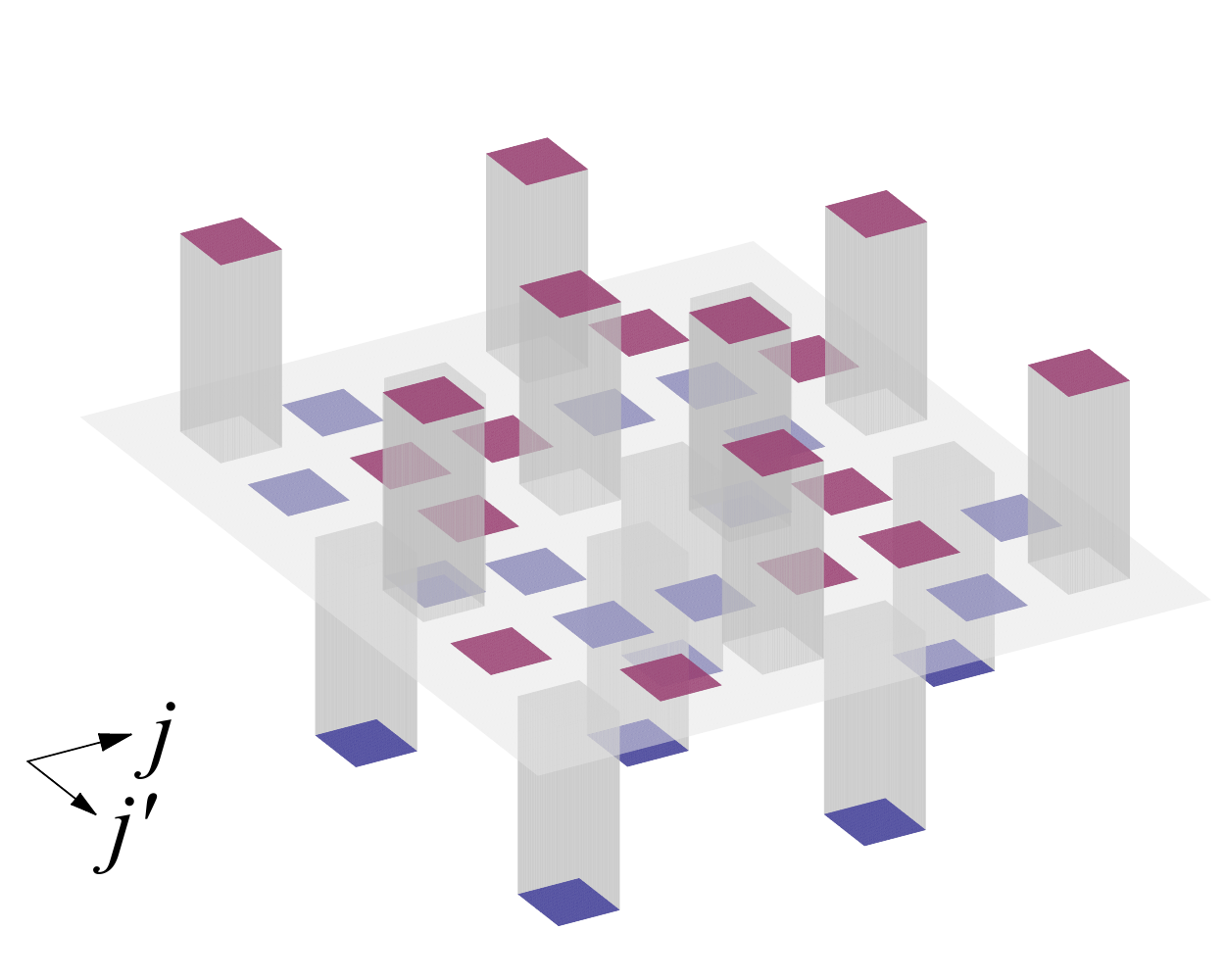}\quad%
 \includegraphics[width=0.33\figurewidth]{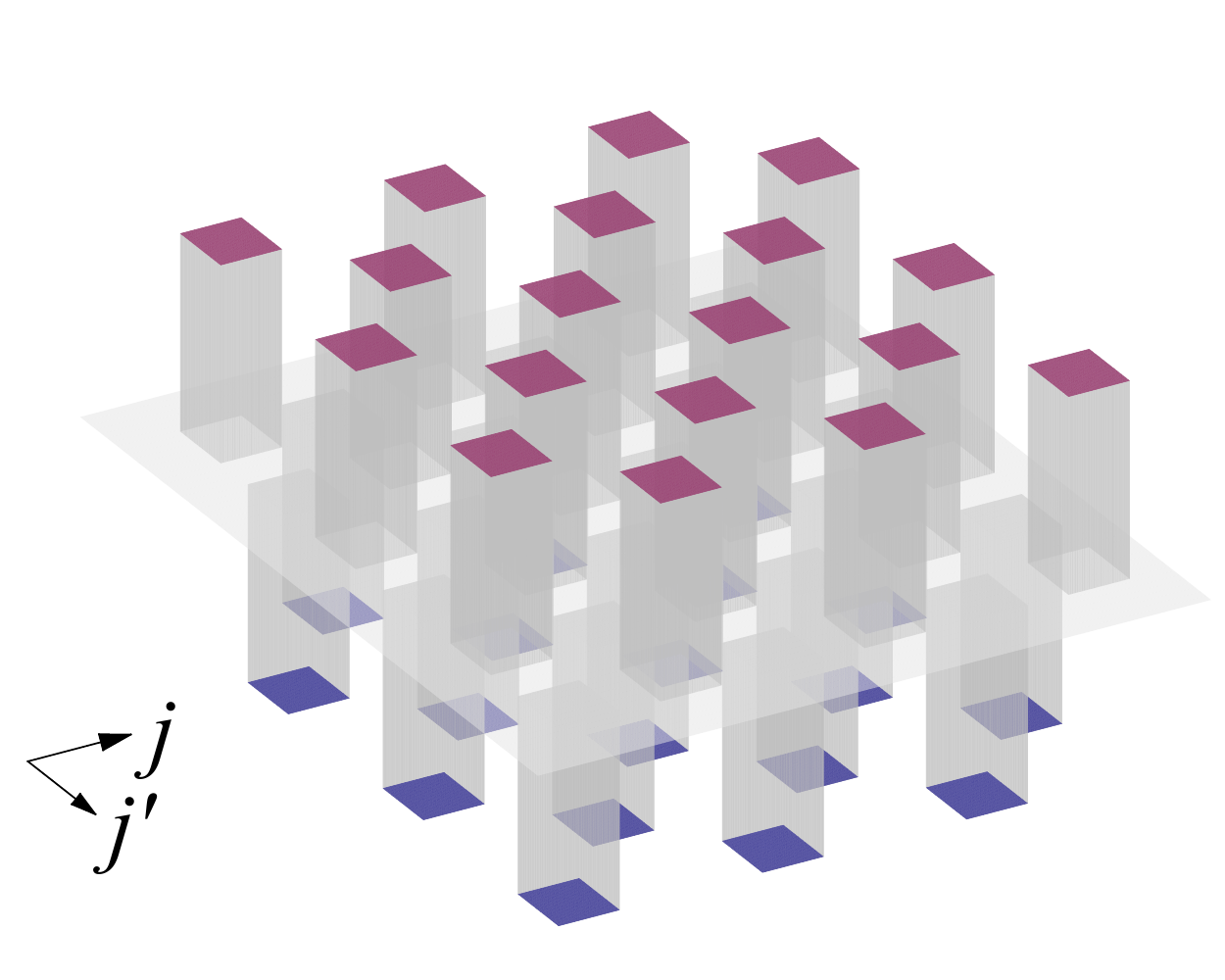}%
 \caption{(Color online) Coupling matrices: $\mat{E}_l=[\epsilon_{l,j}\epsilon_{l,j^\prime}]_{j,j^\prime}$, which determine how excitations flow throughout the array, for $N=6$. Left to right:\ $l=1,2,3$; $l=4$ ($5$) is identical to $l=2$ ($1$). Positive values in magenta, negative in blue.}%
 \label{fig:GMatrices}%
\end{figure}%
\noindent\section{Bad-cavity limit}
\begin{figure}[t]%
 \includegraphics[width=0.9\figurewidth]{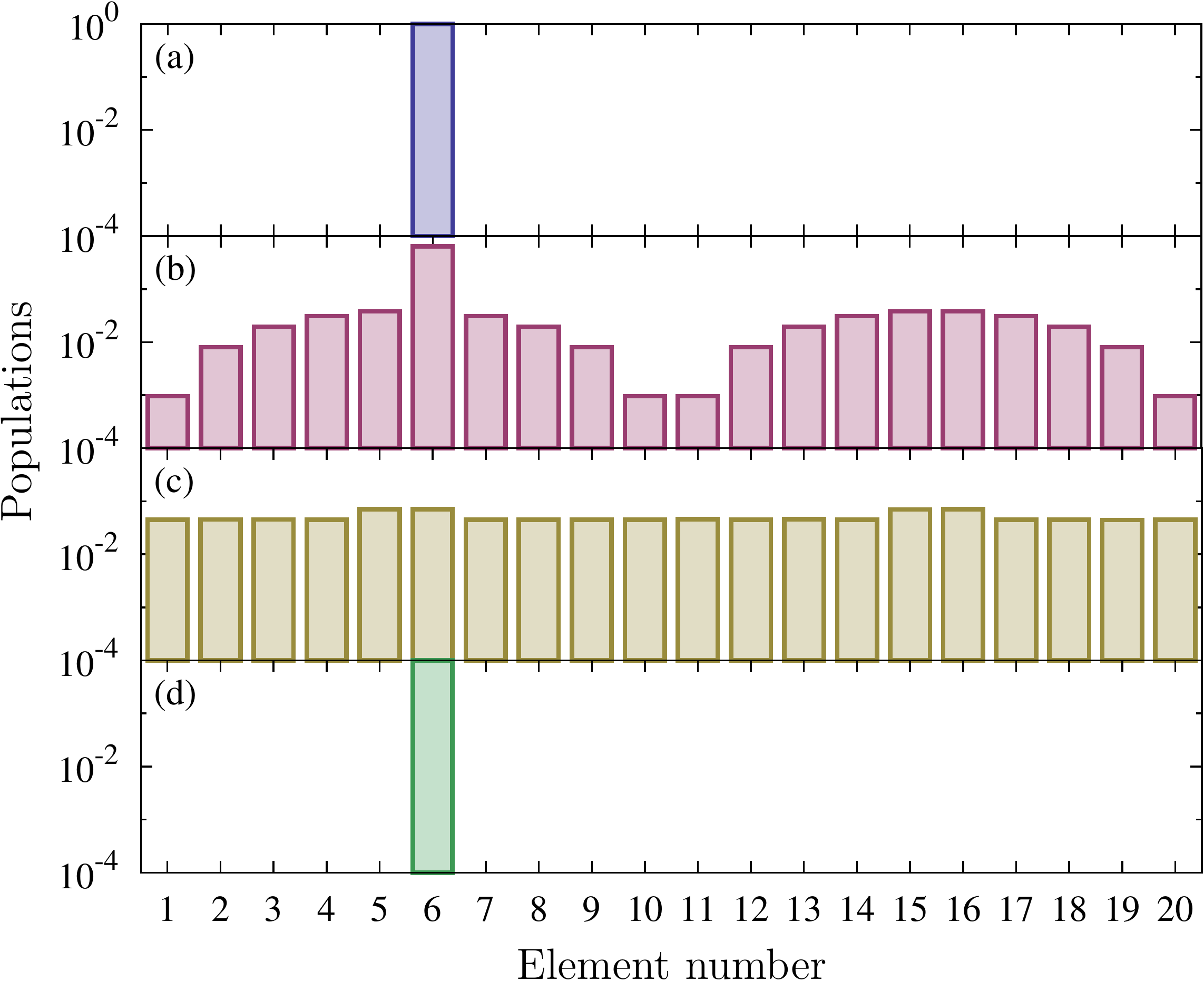}%
\caption{(Color online) A random walk for phonons. Starting from (a)~a coherent state launched from the $6$\textsuperscript{th} element in a $20$-element array, the final populations (b)~in the absence of randomness mimic the vector $\vec{\epsilon}_l$ (we use $l=1$ here). (c)~Randomizing the phase-shifts in the evolution yields a quasi-flat population distribution, whereas (d)~randomizing the transmissivity (parametrized as an angle~\cite{Reck1994}) yields a highly localized distribution. The random angles were drawn from a distribution with mean zero and standard deviation $\pi$; each plot represents an average over $10\,000$ realizations. ($\beta_{l>1}=0$, evolution time $t=\pi/\beta_1$; $\beta_1\ll\omega$ is arbitrary.)}
 \label{fig:Walk}%
\end{figure}

By generalizing the standard procedure~\cite{Gardiner1984}, the optical fields can be eliminated from the dynamics of the optomechanical system provided that $\lvert g_l\rvert\ll\omega\ll\kappa$. This yields the effective linear-coupling Hamiltonian $\hat{H}_\mathrm{eff}=\sum_{l,j,j^\prime}\beta_l\epsilon_{l,j}\epsilon_{l,j^\prime}\hat{b}_j^\dagger\hat{b}_{j^\prime}^{\vphantom{\dagger}}$~(see Appendix) with $\beta_l=2\abs{g_l}^2\Delta_l(\Delta_l^2-\omega^2+\kappa^2)/[(\Delta_l^2-\omega^2-\kappa^2)^2+(2\Delta_l\kappa)^2]$. The matrices $\mat{E}_l=[\epsilon_{l,j}\epsilon_{l,j^\prime}]_{j,j^\prime}$, illustrated in \fref{fig:GMatrices} for $N=6$, and the tuning of $\beta_l$, performed by adjusting $g_l$ and $\Delta_l$, determine how excitations spread through the array. The resulting system is the phononic analog of the random walks explored in Refs.~\cite{Broome2010,Peruzzo2011,Schreiber2011,Sansoni2012,Tillmann2013,Spring2013,Crespi2013,Jeong2013b}.
\begin{figure}[b]%
 \includegraphics[width=0.65\figurewidth]{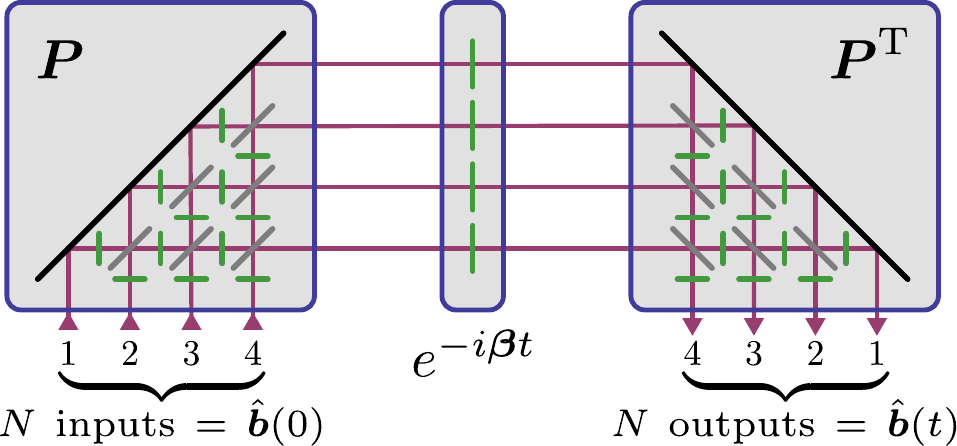}%
\caption{(Color online) Decomposition of the Heisenberg-picture propagator for the vector $\hat{\mat{b}}$~\cite{Reck1994}. $\mat{P}$ is decomposed into beam-splitters (gray; at 45$^\circ$) and phase shifters (green; horizontal and vertical); adding randomness influences the spread of phonons in the array. We show $N=4$.}
 \label{fig:RZ}%
\end{figure}
Using the vectors $\vec{\epsilon}_l$ to build an orthonormal `similarity' matrix $\mat{P}$, we can cast the evolution of the operators describing the mechanical modes as $\expt{\hat{\vec{b}}(t)}=(\mat{P}^\tra e^{-i\mat{\beta}t}\mat{P})\expt{\hat{\vec{b}}(0)}$~(see Appendix), where $\vec{\beta}=(\beta_l\delta_{l,j})_{l,j}$. In a similar fashion to Ref.~\cite{Reck1994}, the matrix $\mat{P}$ can be decomposed into linear optics components (cf.\ \fref{fig:RZ}), allowing a general and physically transparent description of the dynamics, and illustrating the way phonons flow through the array.\\
\Fref{fig:Walk} illustrates a situation where phonons are initially prepared in a coherent state localized at one element of the optomechanical array [panel (a)]. As expected, panel~(b) shows that $\hat{H}_\mathrm{eff}$ imposes a final population distribution with a sinusoidal shape mimicking that of $(\lvert\epsilon_{l,j}\rvert^2)_j$. [It can be demonstrated numerically that if the coherences between the different modes are set to zero after each step in the interferometer, the resulting ``classical'' distribution does not bear any resemblance to $(\lvert\epsilon_{l,j}\rvert^2)_j$.] Furthermore, we find that the population distribution can be modified by a proper design of the effective beam splitters and phase shifters described above~\cite{Jeong2004}. As examples of this flexibility in the manipulation of phonon dynamics, we impose two kinds of randomness on the system:~(i)~a random phase offset to the phase shifters making up $\mat{\beta}$, which can be generated by adding noise to the optical parameters; and (ii)~a randomization of the transmission of the beam-splitters in the decomposition of $\mat{P}$, which corresponds to perturbing $\epsilon_{l,j}$, i.e., changing the properties of the mechanical elements~\cite{Xuereb2013}. In the former case, Fig.~\ref{fig:Walk}(c) shows that averaging over many realizations of random phase distributions yields almost uniform phonon populations. In the latter case, panel~(d) shows that disorder has instead the opposite effect:~the probability distribution collapses into a highly-localized one with significant population only at the element where the excitation was injected. Combinations of these possibilities can be realized, resulting in a flexible control of the type of phonon walk to be implemented.

Let us now explore the flow of heat throughout the array. The Hamiltonian $\hat{H}$ described above is quadratic and therefore preserves the Gaussian nature of any input state of this kind. With this in mind, we constrain the present analysis to the set of Gaussian states. Each of the mechanical elements is coupled to a Markovian bath characterized by a temperature giving rise to a mean number of excitations $\bar{n}_j$ in element $j$. We choose $\bar{n}_j=\bar{n}+\Delta n\,\delta_{j,J}$ for $1\leq J\leq N$. Therefore, each phonon bath has a mean number of excitations $\bar{n}$ except for that of element $J$, which has $\bar{n}+\Delta n$. The heat dynamics in the array is then analyzed by solving the differential equation governing the evolution of the covariance matrix of the $(2N-1)$-partite system~\cite{Ferraro2005}.\\
\begin{figure}[t]%
 \includegraphics[width=0.9\figurewidth]{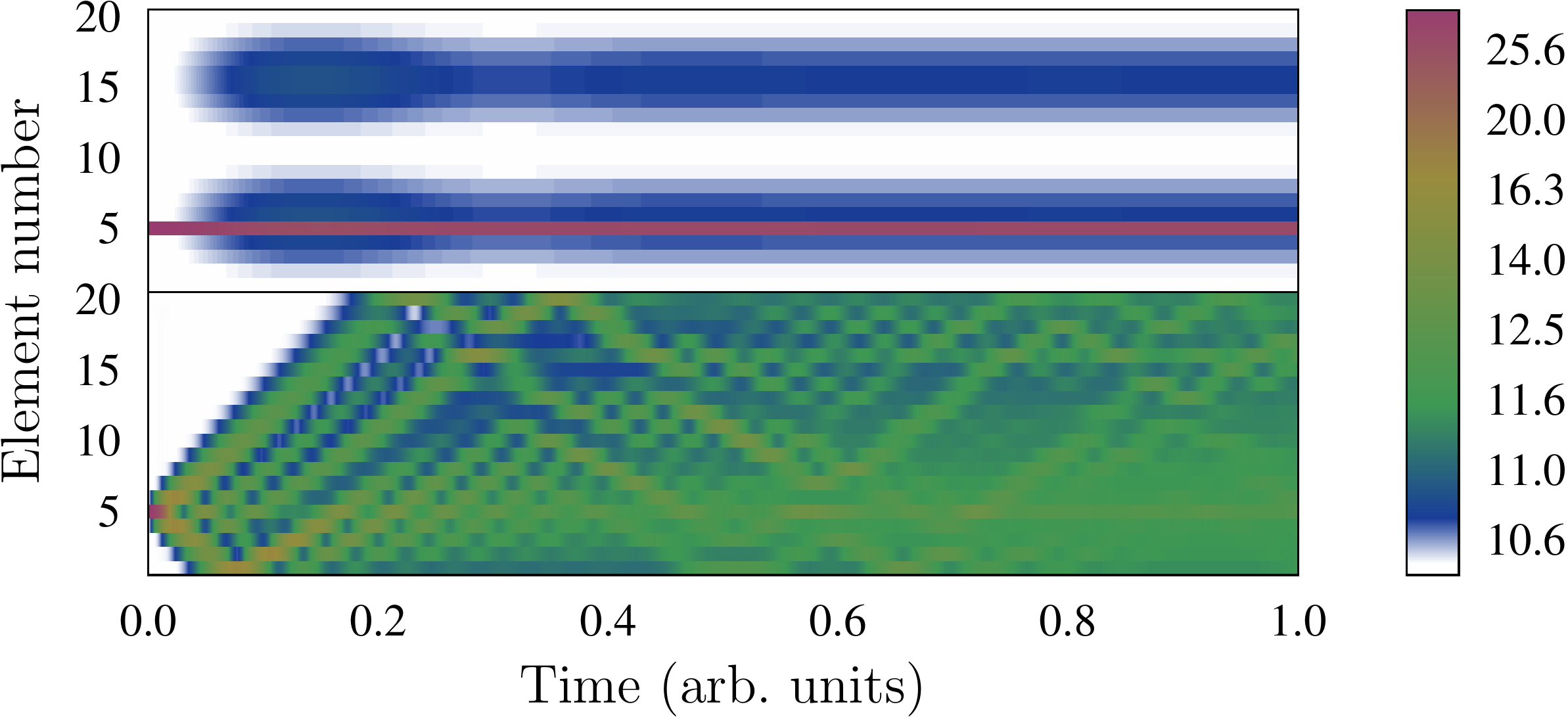}%
 \caption{(Color online) Heat diffusion in optomechanical arrays. Top: We evaluate the mean excitation number on each of $20$ elements as a function of time; initially each element is in thermal equilibrium with the bath that it is in contact with in the absence of light. The optical field causes excitations to flow from the `hotter' element to the rest of the array. In marked contrast with a nearest-neighbor coupling (bottom), the flow does not take place via conduction through adjacent elements, but is mediated by the optical field and occurs to the entire array simultaneously. ($\gamma/\omega=5\times10^{-5}$, $\kappa/\omega=6.4$, $\Delta_l/\omega=-1$, $g_1/\omega=0.3$, $g_{>1}=0$, $\bar{n}=10$, $\Delta n=20$. The nearest-neighbour coupling strength was chosen to be $0.3\,\omega$.)}%
 \label{fig:HeatFlow}%
\end{figure}%
The adiabatic elimination discussed previously yields a system of $N$ harmonic oscillators coupled not only to the aforementioned $N$ thermal baths, but also to $N-1$ shared reservoirs. These reservoirs, which arise through the coupling of each optical field to collective mechanical modes~\cite{Xuereb2013}, complicate the picture and prevent the standard identification of `heat flowing through an element' in the spirit of Ref.~\cite{Asadian2013}. The alternative we explore in this section is to calculate the occupation number of the $N$ mechanical elements and infer from this the effective heat flow through the array.\\
The results, illustrated in \fref{fig:HeatFlow}, exhibit two `non-standard' behaviors that are intimately tied with the properties of the optomechanical system under consideration. First, this system exhibits \emph{static reconfigurability}, i.e., the form of the steady-state phonon population distribution can be chosen by changing which of the optical fields is used to induce the indirect couplings between them. Whilst it is not possible to choose an arbitrary distribution, owing to the symmetry of the effective Hamiltonian, this choice still admits considerable flexibility. Second, the flow of energy from one mechanical element to another takes place indirectly, through the cavity field. It therefore proceeds at a similar rate throughout the entire array, governed not by the distance between the source element $J$ and the element in question but by the coupling constant of the latter to the optical field. A corollary of this is:\ if $\epsilon_{l,j^\prime}=0$ for some $j^\prime$, one can speak of heat flow from the element $J<j^\prime$ to another element $j>j^\prime$ without necessitating any form of heat conduction through element $j^\prime$ itself. This situation occurs, e.g., for $l=1$ and $j^\prime=(N+1)/2$ whenever $N$ is odd~\cite{Xuereb2013}. For even $N$, the elements closest to the center of the array are the least affected. What distinguishes optically-mediated from direct coupling is thus (i) reconfigurability; and (ii) timescales, as excitations flow to every element simultaneously in the optical case, rather than sequentially. These studied interactions enable the study of heat transfer and thermodynamics in non-standard settings~\cite{Dubi2011,Bermudez2013}. The parameters selected for plotting \fref{fig:HeatFlow} were such that for $\Delta n=0$ the steady-state occupation numbers were, to a good approximation, all equal to $\bar{n}$, regardless of the cooling effects of the optomechanical coupling.

\noindent\section{Good-cavity limit}
We now turn to the `good-cavity' regime, for which $\kappa\ll\lvert g_l\rvert\ll\omega$. This allows us to neglect the non-unitary dynamics in an approximate picture if we confine ourselves to times $\ll1/\kappa$. Upon setting $\Delta_l=-\omega$, moving into a rotating frame with respect to the free Hamiltonian (i.e., $\hat{H}$ with $g_l=0\ \forall\ l$), and neglecting rapidly-oscillating terms in the Hamiltonian, we obtain the evolution operator
\begin{equation}
\hat{U}(t)=\exp\bigl[-it{\textstyle\sum_{j,l}\epsilon_{l,j}(g_l^{\vphantom{\ast}}\hat{a}_l^\dagger\hat{b}_j^{\vphantom{\dagger}}+g_l^\ast\hat{a}_l^{\vphantom{\dagger}}\hat{b}_j^\dagger)}\bigr]\,.
\end{equation}
Our interest lies in the coherent shuttling of single excitations around the system. We therefore constrain ourselves to the single-excitation subspace and express the state vector as a $(2N-1)$-dimensional vector with the first $N-1$ (last $N$) entries representing the probability amplitude of the excitation to be found in the respective optical (mechanical) mode. To simplify the notation, let us define the matrix $\mat{\Lambda}=[ig_l^\ast \epsilon_{l,j}]_{l,j}$ ($N$ columns, $N-1$ rows). It can then be shown that the unitary evolution matrix can be written in the block-matrix form~(see Appendix)
\begin{equation}
\label{eq:UMatrix}
\mat{U}(t)=\begin{bmatrix}
\phantom{+}\mat{u}_{11}^{\vphantom{\dagger}} & \mat{u}_{12}^{\vphantom{\dagger}}\\
-\mat{u}_{12}^\dagger & \mat{u}_{22}^{\vphantom{\dagger}}
\end{bmatrix}\,,
\end{equation}
where $\mat{u}_{11}=\cos\bigl(t\sqrt{\mat{\Lambda}\mat{\Lambda}^\dagger}\bigr)$, $\mat{u}_{22}=\cos\bigl(t\sqrt{\mat{\Lambda}^\dagger\mat{\Lambda}}\bigr)$, and $\mat{u}_{12}=-\mat{\Lambda}\sin\bigl(t\sqrt{\mat{\Lambda}^\dagger\mat{\Lambda}}\bigr)\bigl(\sqrt{\mat{\Lambda}^\dagger\mat{\Lambda}}\bigr)^{-1}$~\footnote{This is a multi-mode generalization, restricted to the single-excitation subspace, of the Jaynes--Cummings evolution operator~\cite{Stenholm1973}.}\nocite{Stenholm1973}.\par
In principle, this evolution can even be \emph{dynamically reconfigurable} if we allow for the possibility that the amplitudes $g_l$ of the optical modes can be changed on a timescale $\ll1/\omega$, and therefore significantly shorter than any other timescale of the problem. The implementation of this is discussed in detail in the Appendix; we note that it is crucial that this switching occurs when the mechanical and optical subsystems are uncorrelated and no excitations reside in the optical subsystem. With this in mind, we can therefore string together sequences of $\mat{U}(\bullet)$, between which the amplitudes $g_l$ are changed instantaneously.
\begin{figure}[t]%
 \includegraphics[width=0.95\figurewidth]{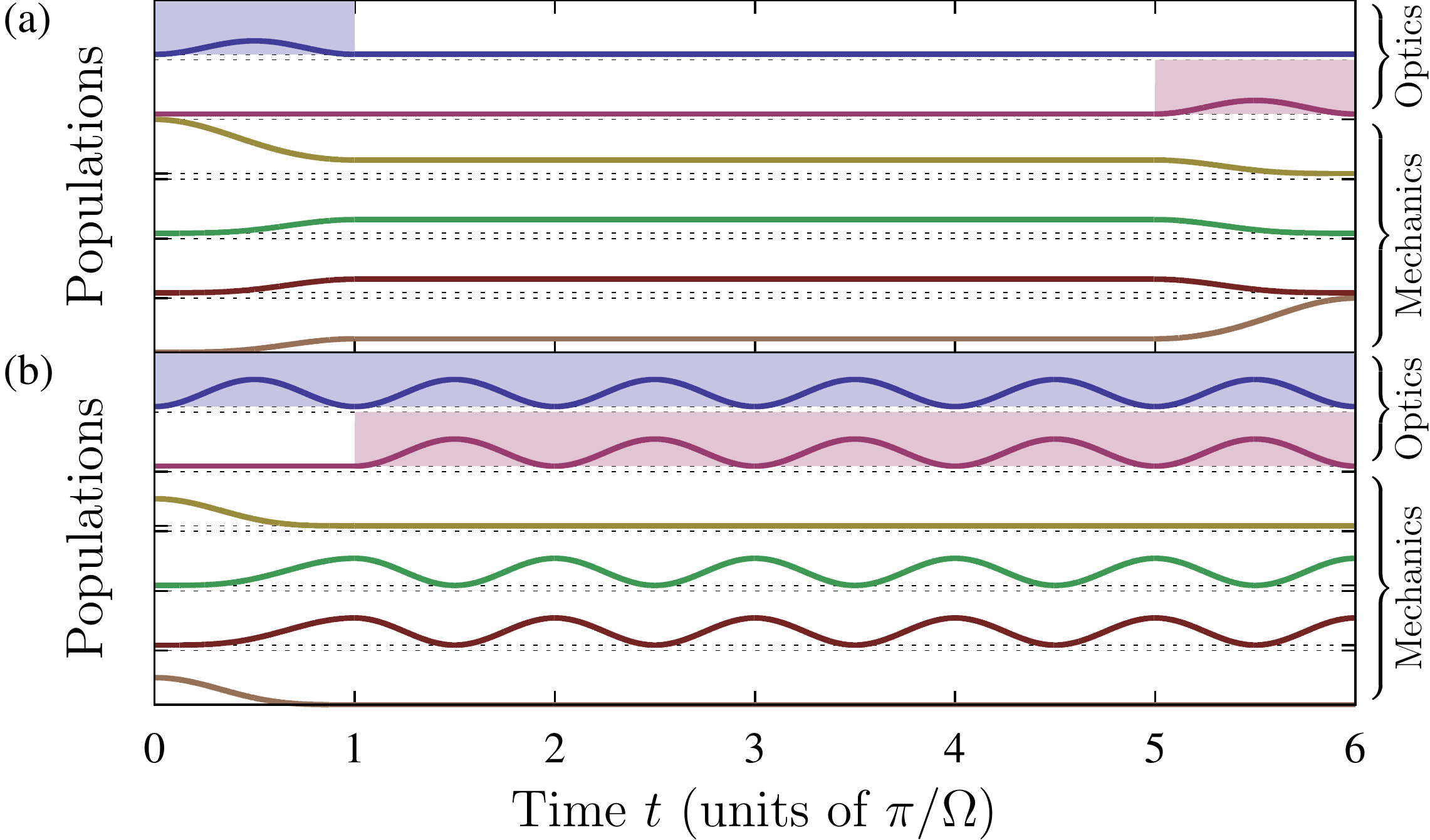}%
\caption{(Color online) Phonon shuttling in an optomechanical array. Shown are (top to bottom) the populations in two optical fields and four mechanical elements, offset for clarity. Shaded regions denote when the mean component of the optical field is nonzero. (a)~`Hold-and-switch' protocol: A phonon on element $1$ is transferred to a superposition and then to element $4$. (b)~A phonon initially in a superposition state between elements $1$ and $4$ is transferred into an excitation shared between the mechanical $2$ and $3$ and the light fields. For $\omega\gg\kappa$ the short-time dynamics is not affected appreciably upon inclusion of dissipation~(see Appendix). (`Off' amplitudes $g_l=0$, `on' amplitudes $g_l=\Omega$; $\Omega\ll\omega$ is arbitrary.)}%
 \label{fig:Shuttling}%
\end{figure}%
The result of this procedure is a set of linear equations that allow us to engineer the route of an excitation through the array. As an example, we illustrate the case for $N=4$, where the fact that $\abs{\epsilon_{l,j}}=\tfrac{1}{2}$ independently of $l$ and $j$ allows for particularly simple protocols to be devised. We demonstrate our ideas by means of the two different examples shown in \fref{fig:Shuttling}: (a)~By switching the amplitudes of two fields, we transport a phonon from mirror $1$ to mirror $4$, and (b)~starting from an initial superposition of the phonon on mirrors $1$ and $4$, we end up with a polariton oscillating between mirrors $2$ and $3$ and the light fields.

\noindent\section{Discussion and outlook}
We have investigated collective dynamics in multi-mode OMS with the goal of simulating many-body effects. The dynamical regimes considered in our analysis showcases distinctive possibilities, ranging from diffusion-like propagation of phononic excitations across the array to the controlled transfer of phonons between targeted elements of the mechanical system. Other regimes of interest could be similarly explored. For instance, operating with blue-detuned cavity fields would allow for investigating collective self-oscillations and synchronization~\cite{Ludwig2012b} in such systems; exploiting the intrinsic nonlinearity of the optomechanical coupling could enable simulation of many-body models (e.g., the Bose--Hubbard Hamiltonian)~\cite{Jacobs2012} or quantum information processing~\cite{Rips2013} with mechanical systems; and using ring cavities would allow exploring geometric phases~\cite{Chesi2014}. Such studies are promising for engineering non-trivial many-body dynamics, a possibility we plan to pursue in future works addressing dissipative quantum state engineering, dynamical phase transitions, and fluctuation theorems of thermodynamics origin~\cite{Mazzola2013,Dorner2013,Mazzola2014}.

\noindent\section{Acknowledgements}
A.X.\ would like to thank C.\ Di Franco for interesting discussions and the Royal Commission for the Exhibition of 1851 for financial support. C.G.\ acknowledges support from the Austrian Science Fund (FWF):\ P24968-N27, and G.P.\ from the European Commission through ERC St-Grant ``COLDSIM'' (No.\ 307688), AFOSR, and UdS Labex. M.P.\ thanks the UK EPSRC for a Career Acceleration Fellowship and a grant awarded under the ``New Directions for Research Leaders" initiative (EP/G004579/1), the Alexander von Humboldt Stiftung, the John Templeton Foundation (grant 43467), and the EC Collaborative project TherMiQ (No.\ 618074). A.D.\ acknowledges funding from the EU (CCQED project), the Institut Francais du Danemark (IFD2013 program) and the Danish Council for Independent Research (Sapere Aude program). Some calculations were carried out using computational facilities funded by the European Regional Development Fund, Project ERDF-080.

\appendix
\section{Choice of working point: Linearization}
Throughout the main text, we worked exclusively in the linearized domain, assuming that the single-photon coupling strength between each of the mechanical elements and the optical fields is small. As is well-known~\cite{Meystre2013,Aspelmeyer2013}, upon application of a strong driving field this means that the effective optomechanical coupling strength is equal to the single-photon coupling strength multiplied by the square-root of the expectation value of the photon number inside the relevant optical mode. This photon number can be chosen arbitrarily through the strength of the driving field and, therefore, so can the effective coupling strength. This means that we do not need to optimize the parameters of the system (in the optical case, this would mean having to carefully choose highly-reflective mirrors and a small cavity) as done when trying to obtain strong single-photon coupling: What we require is simply that the physical configuration is chosen such that the system is in the transmissive configuration defined in Ref.~\cite{Xuereb2012c}. We note, however, that experimental limitations (including absorption and limited laser power) and optomechanical instabilities act to limit the maximum effective coupling strength achievable in any one set-up. Under such conditions, it may still be advantageous to design the system to optimise the single-photon coupling strength.\\
This choice of working point is crucial in distinguishing the family of systems we explore from that proposed in Ref.~\cite{Seok2012} and others in the literature. Indeed, our protocols require:
\begin{enumerate}
\item at least two optical modes interacting with different mechanical oscillators, and
\item that the resulting collective mechanical oscillators being different superpositions of the same physical oscillators.
\end{enumerate}

\section{Adiabatic elimination of the optical field in the optomechanical master equation}
Here we briefly go through the steps for adiabatically eliminating the cavity field from the master equation describing an optomechanical system consisting of one optical field (annihilation operator $\hat{a}$) interacting with one mechanical mode ($\hat{b}$). Our treatment follows that in Ref.~\cite{Jaehne2008}, the Supplementary Information of Ref.~\cite{Stannigel2012}, and \S5.1.2 in Ref.~\cite{Gardiner2004}. We begin by splitting the optomechanical Hamiltonian~\cite{Law1995} that describes linear coupling to the optical field into the free ($\hat{H}_0$) and interaction ($\hat{H}_1$) parts:
\begin{equation}
\hat{H}_0=\omega\hat{b}^\dagger\hat{b}^{\phantom{\dagger}}-\Delta\hat{a}^\dagger\hat{a}\,,\text{\ and\ }\hat{H}_1=-g\bigl(\hat{b}+\hat{b}^\dagger\bigr)\hat{a}^\dagger\hat{a}\,,
\end{equation}
together with the cavity-driving Hamiltonian $\hat{H}_\mathrm{d}=i\sqrt{2\kappa}\alpha_\mathrm{in}\bigl(\hat{a}^\dagger-\hat{a}\bigr)$, where the (coherent) driving field amplitude $\alpha_\mathrm{in}$ is taken to be real for simplicity. To account for non-unitary dissipative processes, we consider the density matrix $\varrho$ for the composite system and introduce the superoperators
\begin{equation}
D[\hat{c}]\varrho:=2\hat{c}\varrho\hat{c}^\dagger-\hat{c}^\dagger\hat{c}\varrho-\varrho\hat{c}^\dagger\hat{c}\,,
\end{equation}
and
\begin{equation}
D_\mathrm{th}[\hat{c}]:=\bigl(n_\mathrm{th}+1\bigr)D[\hat{c}]+n_\mathrm{th}D[\hat{c}^\dagger]\,,
\end{equation}
where $n_\mathrm{th}$ represents the mean number of excitations in the bath that $\hat{c}$ is coupled to. Given the decay rates $\kappa$ and $\gamma$ for $\hat{a}$ and $\hat{b}$, respectively, we can now write the master equation of our system as
\begin{equation}
\dot{\varrho}=-i\bigl[\hat{H}_0+\hat{H}_1+\hat{H}_\mathrm{d},\varrho\bigr]+\kappa D[\hat{a}]\varrho+\gamma D_\mathrm{th}\!\bigl[\hat{b}\bigr]\varrho\,.
\end{equation}
Our first step is to define two unitary transformations $\hat{U}_a$ and $\hat{U}_b$, which act on $\hat{a}$ and $\hat{b}$, respectively, via the relations
\begin{equation}
\hat{U}_a^{\phantom{\dagger}}\,\hat{a}\,\hat{U}_a^\dagger=\hat{a}+\alpha\,\text{\ and\ }\hat{U}_b^{\phantom{\dagger}}\,\hat{b}\,\hat{U}_b^\dagger=\hat{b}+\beta\,.
\end{equation}
These are used to transform our master equation to (i)~eliminate $\hat{H}_\mathrm{d}$, and (ii)~shift $\hat{a}$ and $\hat{b}$ to have zero mean value. The resulting Hamiltonian can be truncated to be quadratic in the operators, and can be written
\begin{equation}
\omega\hat{b}^\dagger\hat{b}-\Delta\hat{a}^\dagger\hat{a}-g\bigl(\hat{b}+\hat{b}^\dagger\bigr)\bigl(\alpha^\ast\hat{a}+\alpha\hat{a}^\dagger\bigr)\,,
\end{equation}
after a redefinition of $\Delta$ to absorb $\beta$. Let us call these three terms $\hat{H}_\mathrm{s}$, $\hat{H}_\mathrm{b}$, and $\hat{H}_\mathrm{i}$, respectively. At this point we define the system (`s'), bath (`b'), and interaction (`i') Liouvillians as follows:
\begin{align}
\mathcal{L}_\mathrm{s}\varrho&:=-i\bigl[\hat{H}_\mathrm{s},\varrho\bigr]+\gamma D_\mathrm{th}\!\bigl[\hat{b}\bigr]\varrho\,,\\
\mathcal{L}_\mathrm{b}\varrho&:=-i\bigl[\hat{H}_\mathrm{b},\varrho\bigr]+\gamma D[\hat{a}]\varrho\,,
\end{align}
and
\begin{equation}
\mathcal{L}_\mathrm{i}\varrho:=-i\bigl[\hat{H}_\mathrm{i},\varrho\bigr]\,,
\end{equation}
and our master equation reads $\dot{\varrho}=\bigl(\mathcal{L}_\mathrm{s}+\mathcal{L}_\mathrm{b}+\mathcal{L}_\mathrm{i}\bigr)\varrho$. We are interested in an equation of motion for the density matrix with $\hat{a}$ eliminated from it. Our treatment will only be valid to lowest order in $g$. We define an operator $\mathbb{P}$ such that
\begin{equation}
\mathbb{P}\varrho:=\rho_\mathrm{c}\otimes\tr_\mathrm{c}\varrho\,,
\end{equation}
which projects $\varrho$ onto a tensor product of the vacuum state for the cavity ($\rho_\mathrm{c}$), which is the steady-state solution for the bath master equation, and the reduced density matrix where the cavity field has been traced out. It is easy to see that $\mathbb{P}$ is a projection operator ($\mathbb{P}^2=\mathbb{P}$) and that, if we define $\mathbb{I}$ as the identity operator, so is $\mathbb{Q}=\mathbb{I}-\mathbb{P}$. Two properties of these operators are particularly useful to us~\cite[\S5.2.1]{Gardiner2004}:
\begin{align}
\mathbb{P}\mathcal{L}_\mathrm{i}\mathbb{P}&=0\,,\text{\ and}\\
\mathbb{Q}\bigl(\mathcal{L}_\mathrm{s}+\mathcal{L}_\mathrm{b}\bigr)&=\bigl(\mathcal{L}_\mathrm{s}+\mathcal{L}_\mathrm{b}\bigr)\mathbb{Q}\,.
\end{align}
By projecting our master equation, we therefore obtain
\begin{align}
\mathbb{P}\dot{\varrho}&=\mathbb{P}\bigl(\mathcal{L}_\mathrm{s}+\mathcal{L}_\mathrm{b}\bigr)\mathbb{P}\varrho+\mathbb{P}\mathcal{L}_\mathrm{i}\mathbb{Q}\varrho\,,\text{\ and}\\
\mathbb{Q}\dot{\varrho}&=\mathbb{Q}\bigl(\mathcal{L}_\mathrm{s}+\mathcal{L}_\mathrm{b}+\mathcal{L}_\mathrm{i}\bigr)\mathbb{Q}\varrho+\mathbb{Q}\mathcal{L}_\mathrm{i}\mathbb{P}\varrho\,.
\end{align}
The adiabatic (`weak coupling', in the sense of being lowest order in $g$) approximation allows us to formally solve the second equation:
\begin{multline}
\mathbb{Q}\varrho(t)=e^{\mathbb{Q}(\mathcal{L}_\mathrm{s}+\mathcal{L}_\mathrm{b}+\mathcal{L}_\mathrm{i})(t-t_0)}\mathbb{Q}\varrho(t_0)\\
+\int_{t_0}^t\rmd\tau\,e^{\mathbb{Q}(\mathcal{L}_\mathrm{s}+\mathcal{L}_\mathrm{b}+\mathcal{L}_\mathrm{i})(t-\tau)}\mathbb{Q}\mathcal{L}_\mathrm{i}\mathbb{P}\varrho(\tau)\,.
\end{multline}
We now take the initial time $t_0\to-\infty$ (which allows us to ignore the first term) and substitute, in the integrand, the zero-order approximation
\begin{equation}
\mathbb{P}\varrho(\tau)=e^{\mathbb{P}(\mathcal{L}_\mathrm{s}+\mathcal{L}_\mathrm{b})(\tau-t)}\mathbb{P}\varrho(t)\,.
\end{equation}
The $\mathcal{L}_\mathrm{i}$ in the exponent can be ignored to this order of approximation in $g$:
\begin{align}
\mathbb{P}\dot{\varrho}&\approx\mathbb{P}\bigl(\mathcal{L}_\mathrm{s}+\mathcal{L}_\mathrm{b}\bigr)\mathbb{P}\varrho\nonumber\\
&\quad+\int_0^\infty\rmd\tau\,\mathbb{P}\mathcal{L}_\mathrm{i}e^{\mathbb{Q}(\mathcal{L}_\mathrm{s}+\mathcal{L}_\mathrm{b})\tau}\mathbb{Q}\mathcal{L}_\mathrm{i}e^{-\mathbb{P}(\mathcal{L}_\mathrm{s}+\mathcal{L}_\mathrm{b})\tau}\mathbb{P}\varrho(t)\,,
\end{align}
or, since $\mathbb{Q}$ commutes with $\mathcal{L}_\mathrm{s}+\mathcal{L}_\mathrm{b}$,
\begin{align}
\mathbb{P}\dot{\varrho}&\approx\mathbb{P}\bigl(\mathcal{L}_\mathrm{s}+\mathcal{L}_\mathrm{b}\bigr)\mathbb{P}\varrho\nonumber\\
&\quad+\int_0^\infty\rmd\tau\,\mathbb{P}\mathcal{L}_\mathrm{i}\mathbb{Q}e^{(\mathcal{L}_\mathrm{s}+\mathcal{L}_\mathrm{b})\tau}\mathcal{L}_\mathrm{i}e^{-\mathbb{P}(\mathcal{L}_\mathrm{s}+\mathcal{L}_\mathrm{b})\tau}\bigl(\rho_\mathrm{c}\otimes\tr_\mathrm{c}\varrho\bigr)\\
&=\mathbb{P}\bigl(\mathcal{L}_\mathrm{s}+\mathcal{L}_\mathrm{b}\bigr)\mathbb{P}\varrho\nonumber\\
&\quad+\int_0^\infty\rmd\tau\,\mathbb{P}\mathcal{L}_\mathrm{i}\mathbb{Q}e^{(\mathcal{L}_\mathrm{s}+\mathcal{L}_\mathrm{b})\tau}\mathcal{L}_\mathrm{i}\bigl(\rho_\mathrm{c}\otimes e^{-\mathcal{L}_\mathrm{s}\tau}\tr_\mathrm{c}\varrho\bigr)\,.
\end{align}
Finally, we trace over the cavity field to obtain the approximate master equation for the reduced density matrix $\varrho:=\tr_\mathrm{c}\rho$,
\begin{equation}
\dot{\rho}=\mathcal{L}_\mathrm{s}\rho_\mathrm{b}+\tr_\mathrm{c}\int_0^\infty\rmd\tau\,\mathcal{L}_\mathrm{i}\mathbb{Q}e^{(\mathcal{L}_\mathrm{s}+\mathcal{L}_\mathrm{b})\tau}\mathcal{L}_\mathrm{i}\bigl(\rho_\mathrm{c}\otimes e^{-\mathcal{L}_\mathrm{s}\tau}\rho\bigr)\,.
\end{equation}
Substitution of $\mathcal{L}_\mathrm{i}$ into this expression and application of the rotating-wave approximation, after some algebra, give the final result
\begin{align}
\dot{\rho}&=\mathcal{L}_\mathrm{s}\rho-ig^2\lvert\alpha\rvert^2\im{S(\omega)+S(-\omega)}\bigl[\hat{b}^\dagger\hat{b},\rho\bigr]\nonumber\\
&\quad+g^2\lvert\alpha\rvert^2\bigl\{\re{S(\omega)}D\!\bigl[\hat{b}\bigr]\rho+\re{S(-\omega)}D\!\bigl[\hat{b}^\dagger\bigr]\rho\bigr\}\,,
\end{align}
where the spectral density of the cavity field to zeroth order in $g$ is defined by
\begin{align}
S(\omega)&:=\int_0^\infty\rmd\tau\,e^{i\omega\tau}\langle\hat{a}(t+\tau)\hat{a}^\dagger(t)\rangle\nonumber\\
&=\int_0^\infty\rmd\tau\,e^{[i(\Delta+\omega)-\kappa]\tau}=-\frac{1}{i(\Delta+\omega)-\kappa}\,.
\end{align}
We can therefore identify an effective Hamiltonian
\begin{align}
\hat{H}_\mathrm{eff}&=\hat{H}_\mathrm{s}+g^2\lvert\alpha\rvert^2\im{S(\omega)+S(-\omega)}\hat{b}^\dagger\hat{b}\nonumber\\
&=\hat{H}_\mathrm{s}+\frac{2g^2\lvert\alpha\rvert^2\Delta\bigl(\Delta^2-\omega^2+\kappa^2\bigr)}{\bigl(\Delta^2-\omega^2-\kappa^2\bigr)^2+\bigl(2\Delta\kappa\bigr)^2}\hat{b}^\dagger\hat{b}\,,
\end{align}
as well as the effective cooling and heating Liouvillians
\begin{equation}
\mathcal{L}_\mathrm{cool}:=\Biggl[\bigl(n_\mathrm{th}+1\bigr)\gamma+\frac{g^2\lvert\alpha\rvert^2\kappa}{\bigl(\Delta+\omega\bigr)^2+\kappa^2}\Biggr]D\!\bigl[\hat{b}\bigr]\,,
\end{equation}
and
\begin{equation}
\mathcal{L}_\mathrm{heat}:=\Biggl[n_\mathrm{th}\gamma+\frac{g^2\lvert\alpha\rvert^2\kappa}{\bigl(\Delta-\omega\bigr)^2+\kappa^2}\Biggr]D\!\bigl[\hat{b}^\dagger\bigr]\,.
\end{equation}
This calculation is readily generalized to many fields at different frequencies, which do not interfere, and multiple mechanical elements.

\section{Basis vectors diagonalizing the Hamiltonian}
Let us start from the effective Hamiltonian
\begin{equation}
\hat{H}_\mathrm{eff}=\sum_{l,j,j^\prime}\beta_l\epsilon_{l,j}\epsilon_{l,j^\prime}\hat{b}_j^\dagger\hat{b}_{j^\prime}^{\phantom{\dagger}}\,.
\end{equation}
As shown in previous work~\cite{Xuereb2013}, we note that the vectors $\vec{\epsilon}_l$ and $\vec{\epsilon}_{N-l}$ are identical for symmetry reasons. Therefore, the set of vectors $\{\vec{\epsilon}_l\}$ does not span the $N$-dimensional space necessary to describe the $N$ mechanical oscillators. Indeed, in general this set has $l_0:=\ceil[(N-1)/2]$ linearly-independent vectors. For this reason, we can absorb the $\beta_{l>l_0}$ into the $\beta_{l\leq l_0}$, and set $\beta_{l>l_0}=0$. To proceed, let us complete basis by adding necessary unit vectors, forming the set of orthonormal basis vectors $\{\vec{p}_l\}$, where $\vec{p}_l=\vec{\epsilon}_l$ for $l\leq l_0$. The choice of these vectors is completely arbitrary so long as $\{\vec{p}_l\}$ forms an orthonormal basis for the space. For the most part, different choices of the vectors with $l>l_0$ lead to the same dynamics. However, this does not hold true when we add randomness to the system (as we did to generate Fig.\ 4 of the main text).\par
Let us now define the matrix $\mat{P}$ whose rows are the $\vec{p}_l$, as well as the vector of operators $\hat{\vec{b}}=(\hat{b}_j)_j$ and the diagonal matrix $\mat{\beta}=\diag(\beta_1,\beta_2,\dots,\beta_{l_0},0,0,\dots)$. This leads us to a simple diagonal decomposition of the Hamiltonian: $\hat{H}_\mathrm{eff}=\hat{\vec{b}}^\dagger(\mat{P}^\tra\cdot\mat{\beta}\cdot\mat{P})\hat{\vec{b}}$. In the Heisenberg picture, we therefore simply have $\expt{\hat{\vec{b}}(t)}=(\mat{P}^\tra\cdot e^{-i\mat{\beta}t}\cdot\mat{P})\expt{\hat{\vec{b}}(0)}$.

\section{Choice of parameters for Fig.\ 5}
Standard optomechanical theory applies to the system in Fig.\ 5 in the main text. This means that the light field can itself create or destroy collective mechanical excitations, even if all the mechanical baths are at the same temperature. To present the clearest case, we chose the parameters for this figure to nullify this optical effect: If the baths are held at the same temperature, the resulting populations would then be the same for all the oscillators.
For clarity, only one optical mode was activated (i.e., each $g_l$ was set to zero except for one). We chose to use the $l=1$ mode because it has the longest spatial period and is therefore clearest to depict in figures, but any other mode would have done; we chose $g_1/\omega=0.3$, but any other value would have given rise to qualitatively similar data. Next, $\gamma$ was selected such that the resulting quality factor of the mechanical oscillator, $20\,000$, well within experimental reach~\cite{Karuza2013b}. Following this, $\Delta$ was chosen to lie on the red sideband of the oscillator. Finally, $\kappa$ was selected to minimize the optical-only effect.

\section{Evolution operator matrix}
Suppose we have $N$ mechanical elements and $N-1$ optical fields. Working in the single-phonon subspace, we will work with kets of the form $\ket{\boldsymbol{1}_l,\boldsymbol{0}}$ and $\ket{\boldsymbol{0},\boldsymbol{1}_j}$, where $1\leq l\leq N-1$ and $1\leq j\leq N$. The former indicates a photon in field $l$, the latter a phonon on element $j$. Our free Hamiltonian is
\begin{equation}
\hat{H}_0=-\sum_l\Delta_l\hat{a}_l^\dagger\hat{a}_l+\sum_j\omega_j\hat{b}_j^\dagger\hat{b}_j\,,
\end{equation}
and the interaction Hamiltonian
\begin{equation}
\hat{H}_1=\sum_{l,j}\epsilon_{l,j}(g_l^\ast\hat{a}_l+g_l\hat{a}_l^\dagger)(\hat{b}_j+\hat{b}_j^\dagger)\,.
\end{equation}
$\Delta_l$ is the detuning of the $l$\textsuperscript{th} optical field, $\omega_j$ the mechanical frequency of the $j$\textsuperscript{th} mechanical oscillator, $\epsilon_{l,j}\in\mathbb{R}$ the optomechanical coupling frequency between the $l$\textsuperscript{th} field and $j$\textsuperscript{th} oscillator, and $g_l$ the c-number component of the $l$\textsuperscript{th} field. In the interaction picture with respect to $\hat{H}_0$, and taking $-\Delta_l=\omega_j=:\omega\ \forall\ l,j$ we obtain the Hamiltonian
\begin{align}
\hat{H}^\prime(t)&:=e^{i\hat{H}_0t}\hat{H}_1e^{-i\hat{H}_0t}\\
&=\sum_{ij}\epsilon_{l,j}[e^{i(\Delta_l+\omega)t}g_l^\ast\hat{a}_l^{\vphantom{\dagger}}\hat{b}_j^\dagger+e^{-i(\Delta_l+\omega)t}g_l\hat{a}_l^\dagger\hat{b}_j]\,.
\end{align}
We now define the evolution operator
\begin{align}
\hat{U}(t)&:=\exp\biggl[-i\int_0^t\rmd\tau\,\hat{H}^\prime(\tau)\biggr]\\
&=\exp\biggl[-it\sum_{l,j}\epsilon_{l,j}\bigl(g_l^\ast\hat{a}_l^{\vphantom{\dagger}}\hat{b}_j^\dagger+\mathrm{H.c.}\bigr)\biggr]\,
\end{align}
noting that this expression may be generalized appropriately for $\Delta_l\neq-\omega$. To aid analysis, define
\begin{align}
\Lambda_{l,j}&:=i\epsilon_{jl}g_l\,,\\
\hat{u}&:=\Lambda_{l,j}^\ast\hat{a}_l^{\vphantom{\dagger}}\hat{b}_j^\dagger-\Lambda_{l,j}\hat{a}_l^\dagger\hat{b}_j\,,
\end{align}
such that $\hat{U}(t)=e^{t\hat{u}}$. Thus, for example,
\begin{align}
\hat{u}\ket{\boldsymbol{0},\boldsymbol{1}_1}&=-\sum_l\Lambda_{l1}\ket{\boldsymbol{1}_l,\boldsymbol{0}}\,,\\
\hat{u}^2\ket{\boldsymbol{0},\boldsymbol{1}_1}&=-\sum_{j^\prime,l}\Lambda_{l1}\Lambda_{l,j^\prime}^\ast\ket{\boldsymbol{0},\boldsymbol{1}_{j^\prime}}\,,\\
\hat{u}^3\ket{\boldsymbol{0},\boldsymbol{1}_1}&=\sum_{j^\prime,l,l^{\prime\prime}}\Lambda_{l1}\Lambda_{l,j^\prime}^\ast\Lambda_{l^{\prime\prime}j^{\prime}}\ket{\boldsymbol{1}_{l^{\prime\prime}},\boldsymbol{0}}\,,\\
\hat{u}^4\ket{\boldsymbol{0},\boldsymbol{1}_1}&=\sum_{j^\prime,l,l^{\prime\prime}}\Lambda_{l1}\Lambda_{l,j^\prime}^\ast\Lambda_{l^{\prime\prime}j^{\prime}}\Lambda_{l^{\prime\prime}j^{\prime\prime\prime}}^\ast\ket{\boldsymbol{0},\boldsymbol{1}_{j^{\prime\prime\prime}}}\,,
\end{align}
etc. Define the matrix ($N$ columns, $N-1$ rows):
\begin{equation}
\mat{\Lambda}:=\bigl[\Lambda_{l,j}\bigr]_{l,j}\,,
\end{equation}
which allows us to compactly write quantities of the type
\begin{equation}
\sum_l\Lambda_{l1}\Lambda_{l,j}^\ast=\bigl(\mat{\Lambda}^\dagger\cdot\mat{\Lambda}\bigr)_{j1}\,.
\end{equation}
Now, with each $\ket{\psi}$ we associate a $(2N-1)$-element vector
\begin{equation}
\ket{\psi}=\begin{pmatrix}
\braket{\boldsymbol{1}_1,\boldsymbol{0}}{\psi}\\
\vdots\\
\braket{\boldsymbol{0},\boldsymbol{1}_1}{\psi}\\
\vdots
\end{pmatrix}\,.
\end{equation}
With this notation, and generalizing the above work, we obtain $\mat{U}(t)$ as in the main text, where we also made use of the definitions
\begin{align}
\cos(\bullet)&=\tfrac{1}{2}\bigl(e^{i\bullet}+e^{-i\bullet}\bigr)\,,\\
\sin(\bullet)&=\tfrac{1}{2i}\bigl(e^{i\bullet}-e^{-i\bullet}\bigr)\,.
\end{align}
When $\mat{\Lambda}^\dagger\cdot\mat{\Lambda}$ is singular, the form of the resulting equation allows us to use the Moore--Penrose pseudoinverse to compute $\mat{U}(t)$.

\begin{figure}[t]%
 \includegraphics[width=0.975\figurewidth]{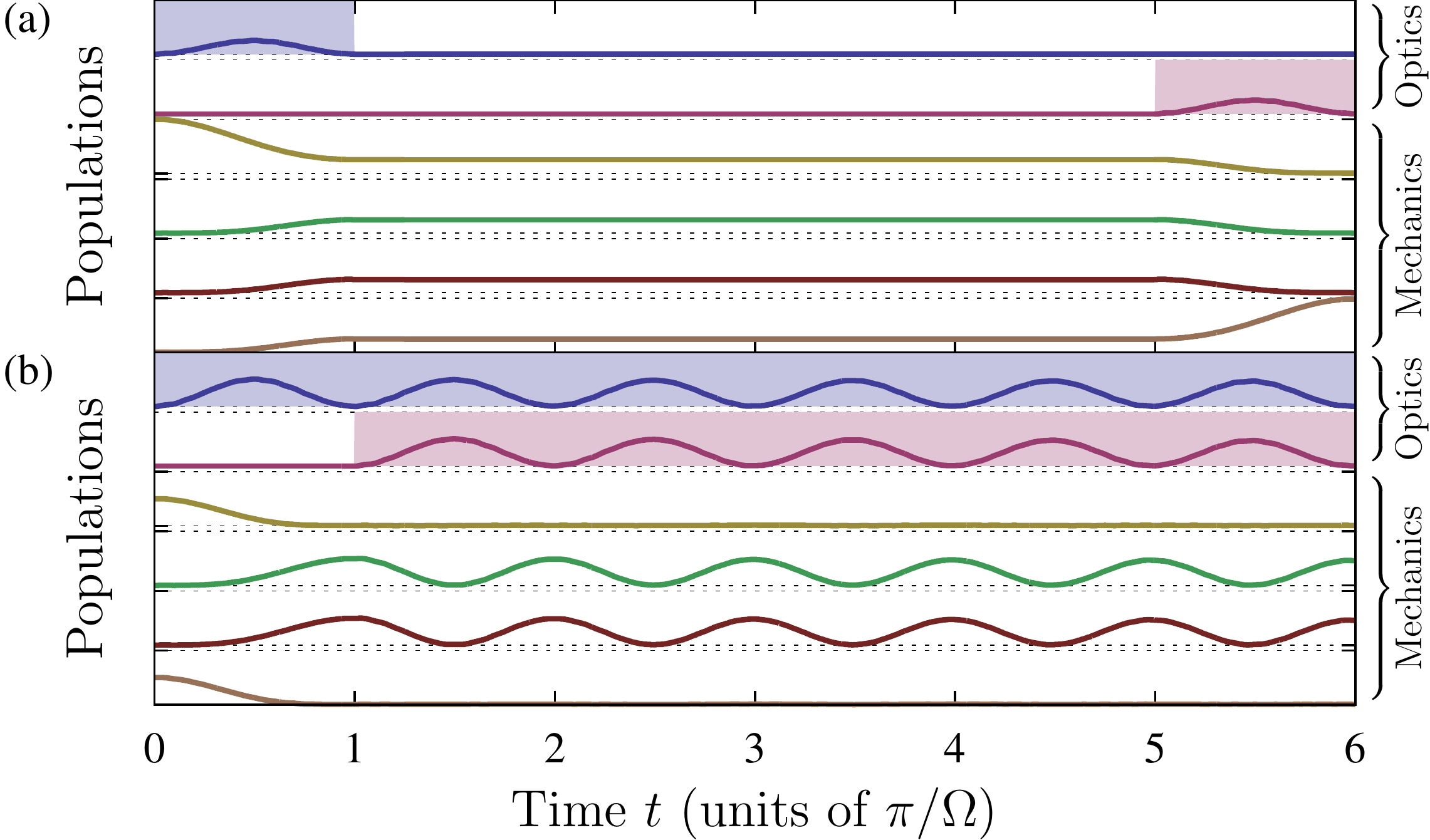}%
\caption{(Color online) Phonon shuttling protocol; same as Fig.\ 6 in the main text but using the full Hamiltonian and including dissipation.}%
 \label{fig:ShuttlingFull}%
\end{figure}%
\section{Non-unitary evolution in the good-cavity limit}
When considering the single-excitation subspace in the good-cavity limit we ignored decay, both optical and mechanical, and took the rotating-wave approximation (RWA). In this section, we present some data to justify these approximations.
First, in \fref{fig:ShuttlingFull}, we simulated the master equation using the full linearized Hamiltonian and the standard dissipation channels. For the chosen parameters ($\kappa=10^{-3}\,\omega$, $\gamma=10^{-6}\,\omega$, $g_1=\Omega=0.1\,\omega$, $g_{>1}=0$), where both the optical and mechanical decay rates are very small, no decay is evident and the unitary approximation, together with the RWA, can be freely used. Breakdown of the RWA could be detected through appearance of higher-frequency oscillations in the populations, which are absent in the presented data. We are presently exploring the positive-detuning case, where the Hamiltonian approximates the ``two-mode squeezing’’ Hamiltonian~\cite{Meystre2013,Aspelmeyer2013}. Under these conditions, one can no longer restrict the analysis to the single-excitation subspace and a concise analysis of the type we explore in the main text is not available.

\section{Dynamical reconfigurability}
In the main text we make use of the fact that the system is \emph{dynamically reconfigurable}. Our protocols necessitate the ability to change the amplitudes $g_l$ of the optical modes on a timescale $\ll1/\omega$, and therefore significantly shorter than any other timescale of the problem. This can be achieved by a rapid change of the cavity linewidth~\cite{Xu2007,Kondo2013,Liu2013}.\\
Furthermore, a straightforward calculation shows that if $\kappa\equiv\kappa(t)$ is a function of time, the input optical field amplitude $\alpha_\mathrm{in}(t)$ can be changed simultaneously to yield an intracavity field amplitude that is constant in time. It is simplest to illustrate this in the case when $\kappa$ changes instantaneously in time, and for a single optical mode; we therefore drop the index labelling the optical modes. Suppose that there exists some time $\tau\gg1/\kappa(0)$ for which $\kappa(t\leq\tau)=\kappa_0$ and $\kappa(t>\tau)=\kappa_1$. The equation of motion for the mean amplitude of the cavity field can be written
\begin{equation}
\dot\alpha(t)=[i\Delta-\kappa(t)]\alpha(t)-\sqrt{2\kappa(t)}\alpha_\mathrm{in}(t)\,,
\end{equation}
where $\Delta$ includes corrections due to the mean amplitude of the mechanical motion. The solution for this equation for a time $t\gg1/\kappa_0$ is
\begin{align}
\alpha(t)&=e^{\int_0^t\rmd t'[i\Delta-\kappa(t')]}\alpha(0)\nonumber\\
&\quad\quad-\int_0^t\rmd t'e^{\int_{t'}^t\rmd t''[i\Delta-\kappa(t'')]}\sqrt{2\kappa(t)}\alpha_\mathrm{in}(t')g\\
&\approx\begin{cases}
-\sqrt{2\kappa_0}\int_0^t\rmd t'e^{(i\Delta-\kappa_0)(t-t')}\alpha_\mathrm{in}(t')\\
-\int_0^t\rmd t'e^{i\Delta(t-t')-[\kappa_0(\tau-t')+\kappa_1(t-\tau)]}\sqrt{2\kappa(t)}\alpha_\mathrm{in}(t')
\end{cases}
\end{align}
where the first line refers to the case where $t\leq\tau$ and the second $t>\tau$. Let us now suppose that $\alpha_\mathrm{in}(t)=\alpha_0$ for $t\leq\tau$ and $\alpha_\mathrm{in}(t)=\alpha_1$ for $t>\tau$. Then,
\begin{align}
\alpha(t)&\approx\begin{cases}
-\sqrt{2\kappa_0}\int_0^t\rmd t'e^{(i\Delta-\kappa_0)(t-t')}\alpha_0\\
-\sqrt{2\kappa_0}e^{(i\Delta-\kappa_1)(t-\tau)}\int_0^\tau\rmd t'e^{(i\Delta-\kappa_0)(\tau-t')}\alpha_0\\
\quad\quad-\sqrt{2\kappa_1}\int_\tau^t\rmd t'e^{i\Delta(t-t')-\kappa_1(t-t')}\alpha_1
\end{cases}\\
&\approx\begin{cases}
\frac{\sqrt{2\kappa_0}\alpha_0}{i\Delta-\kappa_0}\\
\frac{\sqrt{2\kappa_0}\alpha_0}{i\Delta-\kappa_0}e^{(i\Delta-\kappa_1)(t-\tau)}\\
\quad\quad+\frac{\sqrt{2\kappa_1}\alpha_1}{i\Delta-\kappa_1}\bigl[1-e^{(i\Delta-\kappa_1)(t-\tau)}\bigr]
\end{cases}
\end{align}
We now choose $\alpha_1$ such that
\begin{equation}
\frac{\sqrt{2\kappa_1}\alpha_1}{i\Delta-\kappa_1}=\frac{\sqrt{2\kappa_0}\alpha_0}{i\Delta-\kappa_0}\,,
\end{equation}
in which case $\alpha(t)$ does \emph{not} depend on time.
\par
With this in mind, and considering that $g_l\varpropto\alpha$ for the optical mode $l$, one can switch $g_l\to g_l^\prime$ quasi-instantaneously by using the following protocol:
\begin{enumerate}
\item Start from the steady-state at time $t=\tau$, where the intra-cavity mean amplitude is time-indepenent
\item Increase the cavity linewidth, $\kappa\ll\omega\to\kappa^\prime\gg\omega$ quasi-instantaneously, simultaneously adjusting the input field strength according to the above prescription such that the intra-cavity field amplitude stays constant
\item Change the input amplitude; the cavity field re-adjusts over a timescale $\sim1/\kappa^\prime\ll1/\omega$
\item Switch the cavity linewidth back to its original value $\kappa$, once again adjusting the input field amplitude to keep the intra-cavity field amplitude constant
\end{enumerate}
The above protocol `opens up and closes' the cavity, such that any excitations in the optical field are lost. Therefore, we require the excitation to be found entirely in the mechanical subspace during this manipulation, such that the state of the system is not affected by this procedure.
\par
The net effect of this protocol is that for $t>\tau$ the dynamics proceeds as before, albeit with a step-change in the value of $g_l$. With this in mind, we can therefore string together sequences of $\mat{U}(\bullet)$, between which (at times $\tau_1,\tau_2,\dots$) the amplitudes $g_l$ are changed instantaneously, to form the evolution matrix
\begin{equation}
\mat{\mathcal{U}}(t)=\begin{cases}
\mat{U}(t) & : 0\leq t<\tau_1\\
\mat{U}(t-\tau_1)\cdot\mat{U}(\tau_1) & : \tau_1\leq t<\tau_2\\
\ \vdots
\end{cases}\,.
\end{equation}
We note that there are two constraints one must keep in mind when implementing this protocol. First, as already discussed, the switching must occur at a time when the mechanical and optical subsystems are uncorrelated, and when the excitation resides entirely in the mechanical subsystem. Second, the switching time must be much slower than the time-scale set by the inverse of the optical frequency in order to avoid creating photons through mechanisms such as the dynamic Casimir effect.

\section{Practical implementations}
The generic features described in the main text are in principle applicable to a wide range of systems, whether in the optical or the microwave domain. While we used a language pertaining mostly to the optical domain language in the main text, micromechanical elements coupled to superconducting microwave resonator fields represent a very promising system for observing these effects. Indeed, resolved sideband cooling to the ground state and operation in the strong coupling regime~\cite{Teufel2011} have been achieved, and even scalable experiments with multiple elements have been recently carried out~\cite{Massel2012}. Initialization and readout of the mechanics could be performed either by coupling the mechanical elements to additional microwave resonators or to artifical atoms, as has also been demonstrated in Refs.~\cite{OConnell2010,Pirkkalainen2013}, amongst others. In the optical domain potential candidates could be dielectric membranes~\cite{Thompson2008,Karuza2013b,Purdy2013}, forming a periodic array inside an optical resonator, as studied in e.g., Refs.~\cite{Xuereb2012c,Seok2012}. Using parameters close to the experiments of Ref.~\cite{Karuza2013b}, for instance, one could operate in the good cavity regime $\kappa\ll g\ll\omega$ discussed in the paper. Initialization and readout could be performed either optically using standard techniques or electrically using functionalized membranes~\cite{Lee2013b,Adiga2013,Bagci2013,Andrews2013}. In the same fashion toroidal cavities with indentations~\cite{Arbabi2011} could also be used. Another promising candidate for the realization of such a mechanical simulator would be ensembles of cold atoms in optical cavities~\cite{StamperKurn2012,Brahms2012}, for which the required regimes and techniques have all been experimentally demonstrated.

\end{document}